\documentclass{emulateapj}

\usepackage{xcolor}

\defcitealias{2006_Faucher}{FK06}
\shorttitle{Radio efficiency of pulsars}
\shortauthors{Szary et al.}

\begin{document}

    \title{Radio efficiency of pulsars}

    \author{Andrzej Szary\altaffilmark{1,2}, Bing Zhang\altaffilmark{3,2}, George I. Melikidze\altaffilmark{1,4}, Janusz Gil\altaffilmark{1} and Ren-Xin Xu\altaffilmark{2}}

%    \email{aszary@astro.ia.uz.zgora.pl}

    \altaffiltext{1}{Kepler Institute of Astronomy, University of Zielona G\'ora,\\ Lubuska 2, 65-265 Zielona G\'ora, Poland, aszary@astro.ia.uz.zgora.pl}
    \altaffiltext{2}{School of Physics and Kavli Institute for Astronomy and Astrophysics, Peking University, Beijing 100871, China}
    \altaffiltext{3}{Department of Physics and Astronomy, University of Nevada Las Vegas, NV 89154, USA, zhang@physics.unlv.edu}
    %\altaffiltext{4}{Department of Astronomy, School of Physics, Peking University, Beijing 100871, China}
    \altaffiltext{4}{Abastumani Astrophysical Observatory, Ilia State University, 3-5 Cholokashvili Ave., Tbilisi, 0160, Georgia}

    \begin{abstract}
        We investigate radio emission efficiency $\xi$ of pulsars and report a near linear inverse correlation between $\xi$ and the spindown power $\dot E$, as well as a near linear correlation between $\xi$ and pulsar age $\tau$.
        This is a consequence of very weak, if any, dependences of radio luminosity $L$ on pulsar period $P$ and period derivative $\dot{P}$, in contrast to X-ray or $\gamma$-ray emission luminosities.
        The analysis of radio fluxes suggests that these correlations are not due to a selection effect, but are intrinsic to the pulsar radio emission physics.
        We have found that, although with a large variance,  the radio luminosity of pulsars is $\left<L\right>\approx 10^{29} \,{\rm erg/s}$, regardless of the position in the $P-\dot P$ diagram.
        Within such a picture, a model-independent statement can be made that the death line of radio pulsars corresponds to an upper limit in the efficiency of radio emission.
        If we introduce the maximum value for a radio efficiency into Monte Carlo-based population syntheses we can reproduce the observed sample using the random luminosity model.
        The Kolmogorov-Smirnov test on a synthetic flux distribution shows high probability of reproducing the observed distribution.
        Our results suggests that the plasma responsible for generating radio emission is produced under similar conditions regardless of pulsar age, dipolar magnetic field strength, and spin-down rate.
        The magnetic fields near the pulsar surface are likely dominated by crust-anchored magnetic anomalies, which do not significantly differ among pulsars, leading to similar conditions for generating electron-positron pairs necessary to power radio emission.
    \end{abstract}


    \section{Introduction}
        Despite the fact that pulsars were discovered almost half a century ago, the emission mechanism of the pulsed radio emission remains unresolved.
        The challenge lies in the difficulty to identify the correct ``coherent'' mechanism \citep{2006_Melrose} that can power radio emission with similar emission characteristics over about four orders of magnitude in rotation period $P$ and up to eleven orders of magnitude in the rate of increase of pulsar periods $\dot{P} = {\rm d} P / {\rm d} t$.
        In this paper we perform a systematic analysis of radio emission efficiency for the current sample of pulsars\footnote{If not stated otherwise, data presented in this paper are taken from ATNF Pulsar Catalogue {\it http://www.atnf.csiro.au/research/pulsar/psrcat} \citep{2005_Manchester}}, in an effort to study the global radio emission properties and to constrain radio emission mechanism.
        In Section \ref{sec:calc} we present the method to calculate radio luminosity $L$ and radio emission efficiency $\xi$ of pulsars.
        The dependences of both $L$ and $\xi$ on pulsar parameters are presented in Section \ref{sec:res}.
        The results are summarized in Section \ref{sec:sum} along with a discussion on the physical implications of the found correlations.
       \newpage
    \section{Radio luminosity and efficiency} \label{sec:calc}

        \subsection{Integrated radio luminosities}
            For a number of reasons a correct estimate of a pulsar's radio luminosity is difficult \cite[see e.g.][]{2004_Lorimer}.
            A common practice is to assume that the intensity distribution along the observer's line-of-sight cut through the emission region is representative for the entire emission beam.
            Then, the luminosity of a pulsar can be calculated as follows:
            \begin{equation}
                L = \frac{4 \pi d^2}{\delta} \sin^2 \left ( \frac{\rho}{2} \right ) \int_{\nu_{\rm min}}^{\nu_{\rm max}} S_{\rm mean}(\nu) {\rm d}\nu,
            \end{equation}
            where $\delta$ is the pulse duty cycle, $d$ is a distance to the pulsar, $\rho$ is the angular radius of the emitting cone, $\nu_{\rm min}$ and $\nu_{\rm max}$ bracket the radio frequency range in which the pulsar is detected, and $S_{\rm mean}(\nu)$ is the mean flux density measured at a given frequency $\nu$.
            The pulse duty cycle can be calculated using the so called equivalent width $W_{\rm eq}$ (i.e. the width of a top-hat pulse having the same area and peak flux as the true profile) as  $\delta = W_{\rm eq}/P$.
            To derive $\rho$ we have to know the geometry of the pulsar emission beam: $\alpha$ - the inclination angle between the rotation and magnetic axes, $\beta$ - the impact parameter, and $W$ - the observed pulse width.
            These values are usually not known for most pulsars.
            In order to keep the sample of pulsars as large as possible, in this paper we calculate the luminosity $L$ assuming some typical values for all pulsars: $\delta \approx 0.04$, $\rho \approx 6^{\circ}$, $\nu_{\rm min}\approx 10^7 \, {\rm Hz}$, \mbox{$\nu_{\rm max} \approx 10^{11} \, {\rm Hz}$} \cite[see][for more details]{2004_Lorimer}, so that
            \begin{equation}
                L \simeq 7.4 \times 10^{27} \left (\frac{d}{\rm kpc} \right )^2 \left (  \frac{S_{1400}}{\rm mJy} \right ) \,{\rm erg \, s^{-1}}. \label{eq:lum}
            \end{equation}
            Note that here, we neglect any dependence of $\delta^{-1} \sin^2{(\rho/2)}$ on the spin parameters $P$ and $\dot P$.
            In this way we avoid any bias in calculation of $L$ and keep our analysis model-independent (see Section \ref{sec:sum} for discussion).
            Additionally, such an approach allows us to compare the obtained results with X-ray and $\gamma$-ray observations.
            %Since we only care about how $L$ depends on other parameters, the conclusions drawn below are valid as long as the solid angle of radio emission beam does not depend on the spin parameters $P$ and $\dot P$.

        \subsection{Monochromatic luminosities }
            Sometimes it is convenient to use monochromatic luminosities (often also called ``pseudoluminosities'')
            \begin{equation}
                L_{\rm \nu} \equiv S_{\rm \nu} d^2,
            \end{equation}
            to estimate how luminosity varies in frequency. Here $S_{\rm \nu}$ is the mean flux density measured at different frequencies (e.g. $400 \, {\rm MHz}$, $1400 \, {\rm MHz}$ and $2000 \, {\rm MHz}$).
            Note that $L_\nu$ has the similar dependences on $S_{\rm \nu}$ and $d$ as $L$.

        \subsection{Radio emission efficiency}
            The radio emission efficiency of pulsars is defined as the fraction of rotational energy transformed into radio emission, i.e.
            \begin{equation}
                \xi \equiv \frac{L}{\dot{E}},
                \label{eq:eff}
            \end{equation}
            where
            \begin{equation}
                \dot{E} = 4 \pi^2 I \dot{P} P^{-3} \simeq 3.95\times 10^{31} I_{45}\left ( \frac{\dot{P}}{10^{-15}} \right ) \left ( \frac{P}{s} \right )^{-3} \,{\rm erg \, s^{-1}}
            \end{equation}
            is the rate of loss of rotational energy (also called spin-down luminosity), and  $I=10^{45} I_{45} \,{\rm g\, cm^2}$ is the moment of inertia.

            Note that $L$ and $\dot E$ are two independent parameters that come from very different measurements.
            While $L$ is derived from measuring both radio emission flux and distance of a pulsar, $\dot E$ is derived from pulsar timing measurements. The $L$ parameter depends on the unknown pulsar radio emission physics.
            There is no {\em a priori} reason to expect that $\xi$ should depend on $\dot E$ (or other parameters defined by timing parameters) in any way.

    \section{Correlations} \label{sec:res}

            % http://127.0.0.1:8000/database/plots/radio/l_eff_quad
            \begin{figure*}
                \begin{center}
                    \includegraphics{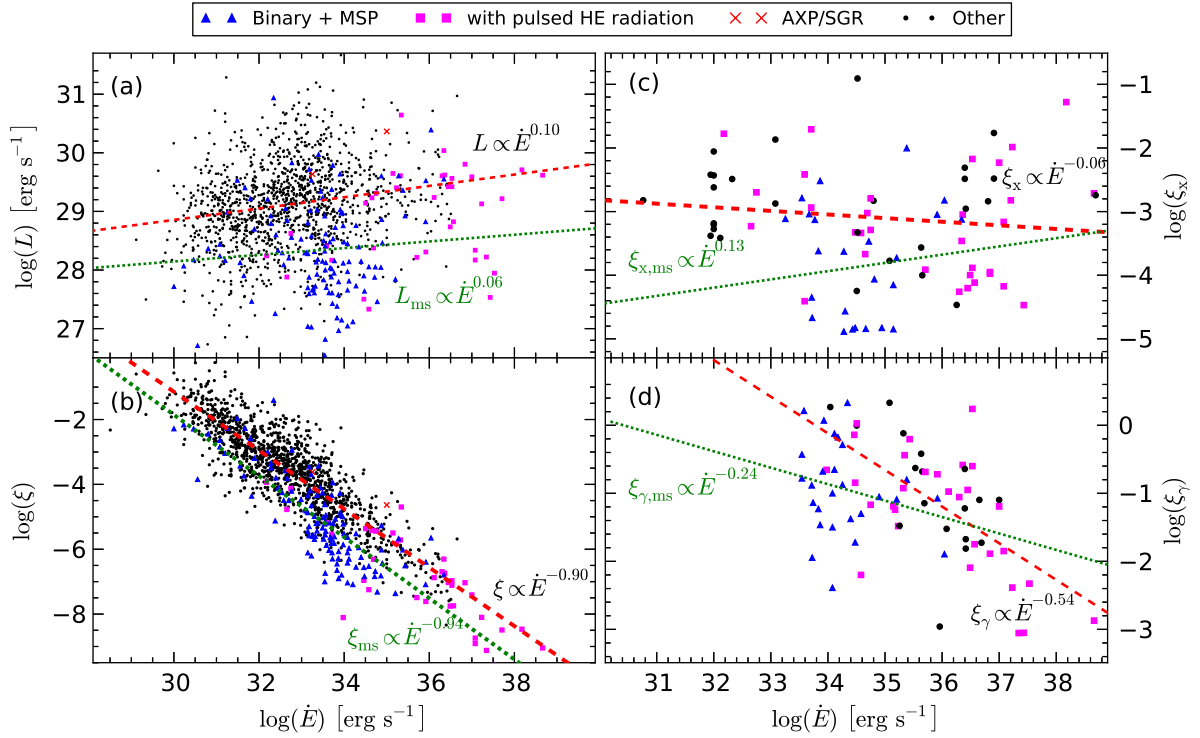}
                \end{center}
                \caption{Dependence of the radio luminosity $L$ (panel a), and radio ($\xi$, panel b), X-ray ($\xi_{\rm x}$, panel c) and $\gamma$-ray ($\xi_{\gamma}$, panel d) efficiencies on the rate of rotational energy losses $\dot{E}$.
                The red dashed lines correspond to the linear fit for normal pulsars, while green dashed line represents the linear fit for binary and millisecond pulsars (blue triangles).
                Pulsars with pulsed high-energetic radiation are marked with magenta squares, whereas anomalous X-ray pulsars (or soft Gamma-ray repeaters) are represented by red crosses.
                The X-ray data were taken from \cite{2013_Szary} for normal pulsars and from \cite{2009_Becker} for millisecond pulsars, while the $\gamma$-ray data were taken from \cite{2013_Fermi}.
                }
                \label{fig:one}
            \end{figure*}

            % http://127.0.0.1:8000/database/plots/radio/radio_xray_gamma
            \begin{figure*}
                \begin{center}
                    \includegraphics{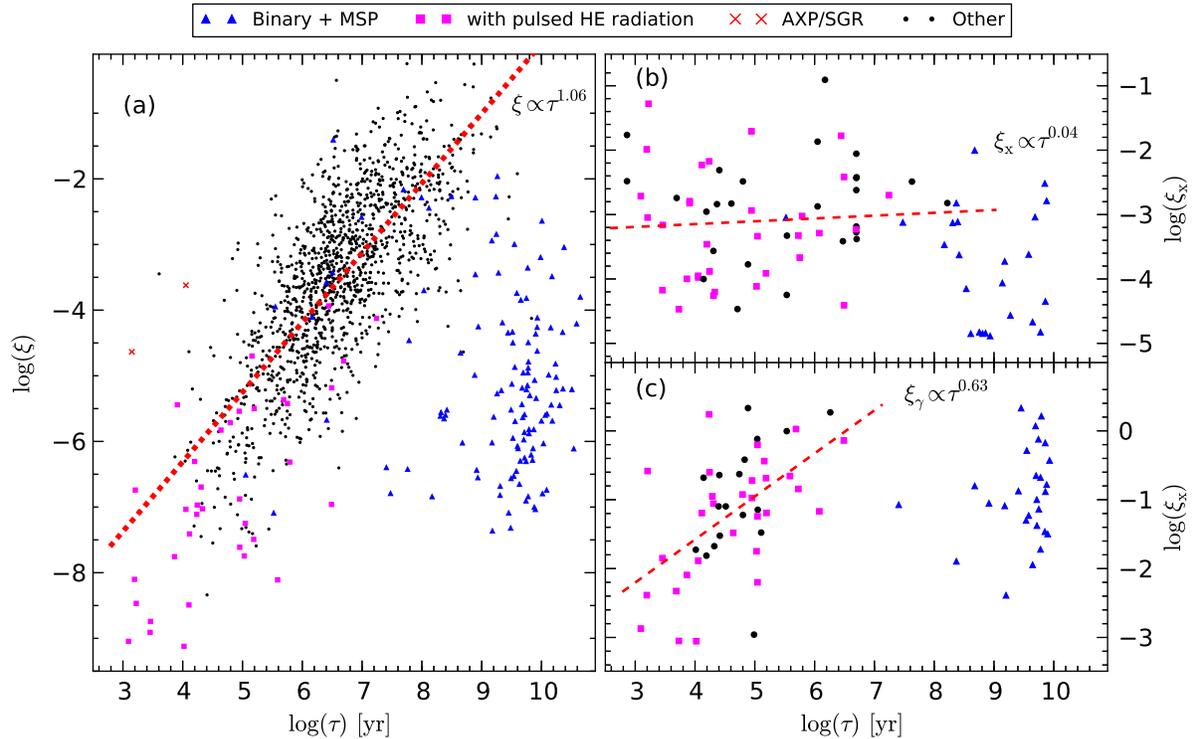}
                \end{center}
                \caption{Dependence of the radio efficiency $\xi$ (panel a), X-ray efficiency $\xi_{\rm x}$ (panel b), and $\gamma$-ray efficiency $\xi_{\gamma}$ (panel c) on pulsar age $\tau$.
                The dashed lines correspond to the linear fit for normal pulsars.
                See Figure \ref{fig:one} for more detailed description.
                }
                \label{fig:two}
            \end{figure*}

        In the following, we investigate how $L$ and $\xi$ are correlated with other pulsar parameters.

        \subsection{Spin-down}
            Figure \ref{fig:one}a shows radio luminosity $L$ as a function of the spin-down power $\dot E$. One immediately sees that there is essentially no dependence between the two parameters, with best fit $L\propto \dot E^{0.10}$ and $L \propto \dot E^{0.06}$ for normal and binary/millisecond pulsars, respectively.
            A similar conclusion is achieved when one replaces $L$ by pseuodoluminosities $L_\nu$.
            The linear fits in different frequencies show that the exponents for $L_\nu-\dot E$ relations are all close to 0, i.e. 0.18, 0.1 and -0.08 for 400, 1400, and 2000 MHz, respectively, with a rough trend that the exponent decreases with increasing frequency.
            Note, however, the pulsar samples at 400 MHz (641 objects) and especially at 2000 MHz (27 objects) are considerably smaller than that at 1400 MHz (1436 objects).

            The weak dependence of $L$ on $\dot E$ suggests a near linear inverse correlation between $\xi$ and $\dot E$.
            Figure \ref{fig:one}b shows such an anti-correlation, with the best linear fits $\xi \propto \dot{E}^{-0.90}$ and $\xi \propto \dot{E}^{-0.94}$ for normal pulsars and binary/millisecond pulsars, respectively.
            Even though the spread in values of both radio luminosity and efficiency is high for a given spin-down luminosity, the $\xi - \dot E$ dependence is clearly visible, with low efficiency (e.g. $\xi=10^{-8}-10^{-5}$) at high spin-down rate (e.g. $\dot{E}=10^{36} \, {\rm erg \, s^{-1}}$) and high efficiency (e.g. $\xi \gtrsim 10^{-3}$) at low spin-down rate (e.g. $\dot{E}=10^{30} \, {\rm erg \, s^{-1}}$).

            Such a near linear inverse correlation between $\xi$ and $\dot E$ is non-trivial.
            For comparison, in panels (c) and (d) in Figure \ref{fig:one} we also show how X-ray and $\gamma$-ray efficiencies depend on $\dot E$.
            It is clearly seen that the X-ray efficiency is essentially independent on $\dot E$, i.e. $\xi_{\rm x} \propto \dot{E}^{-0.08}$ (see Figure \ref{fig:one}c), and the $\gamma$-ray efficiency only weakly depends on $\dot E$, i.e. $\xi_{\gamma} \propto \dot{E}^{-0.5}$ for normal pulsars and as $\xi_{\gamma} \propto \dot{E}^{-0.24}$ for millisecond ones (see Figure \ref{fig:one}d).
            When calculating both the X-ray and $\gamma$-ray efficiencies, we have assumed a same solid angle for all pulsars, which is the same assumption made in calculating radio emission efficiencies.

        \subsection{Age} \label{sec:age}
            From the very beginning of pulsar astronomy it was suggested that radio luminosity of pulsars must decline with age.
            Such an evolution could explain the rapid drop in pulsar distribution around the period of $1 {\rm s}$ \citep{1970_Gunn}.
            As suggested by \cite{1977_Taylor} many more pulsars would be observed if the luminosity were constant.

            Since $\dot E \propto \dot{P} P^{-3}$ and $\tau \propto \dot{P}^{-1} P$, the above negative linear $\xi-\dot E$ correlation would be translated to a positive linear $\xi-\tau$ correlation.
            Figure \ref{fig:two}a shows that such a dependence is indeed there, with a best linear fit (for normal pulsars only) $\xi \propto \dot{\tau}^{1.06}$.
            Notice that millisecond pulsars are excluded in the analysis, since they have experienced recycling spin-up process, so that their $\tau$ is not the characteristic age since birth.

            This interesting relation suggests a surprising result that as a pulsar ages, it somehow transforms its spin-down luminosity more efficiently into radiation.
            Such a relationship is nontrivial, and provides valuable information about the mechanism of radio emission.% (see Section \ref{sec:sum}).

            Again for comparison, we plot $\xi_{\rm x}$ and $\xi_\gamma$ against $\tau$ in Fig. \ref{fig:two}b and Fig. \ref{fig:two}c, respectively. We find unlike radio emission, there is essentially no obvious correlation in X-rays (with $\xi_{\rm x} \propto \tau^{0.11}$), and there is only a weak and rather scattered correlation in $\gamma$-rays (with $\xi_{\rm \gamma} \propto \tau^{0.63}$). This again suggests that the radio emission mechanism is different from those of high-energy radiation.

        \subsection{Selection effect?}
            The lack of pulsars with high-efficiency, high-$\dot E$ and young-age pulsars suggest that there is no significant selection effect for the two reported correlations above.
            One may still suspect that the two correlations in the low-$\dot E$ and old-age regime may be affected by an observational selection effect, which is against the detection of low-$\xi$ pulsars due to the sensitivity limit of radio telescopes.

\begin{figure}[ht]
                \begin{center}
                    \includegraphics[width=7.5cm]{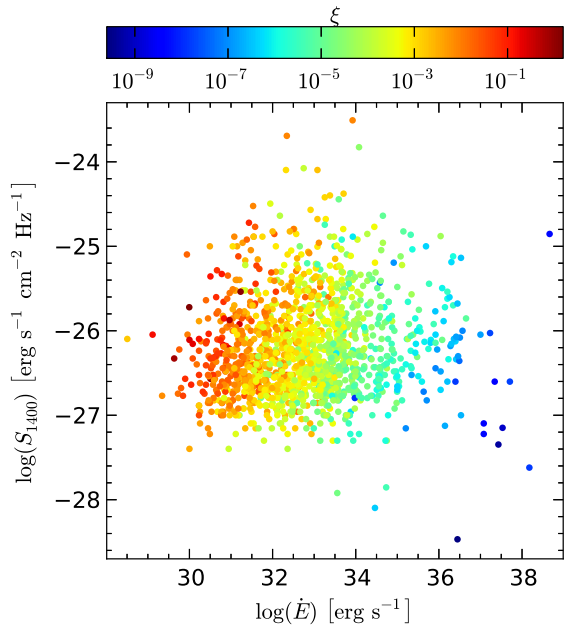}
                \end{center}
                \caption{Dependence of the observed mean flux density measured at $1400 \, {\rm MHz}$ $S_{1400}$ on spin-down luminosity $\dot{E}$. Colors correspond to different values of radio efficiency.}
                \label{fig:four}
            \end{figure}

            Figure \ref{fig:four} presents the dependence of the observed mean flux density $S_{1400}$ measured at $1400 \, {\rm MHz}$  on the spin-down luminosity $\dot{E}$.
            As can be seen from the Figure, the flux distribution of pulsars with relatively low $\dot E$ is not significantly different from the flux distribution of pulsars with relatively high $\dot E$.
            There is no significant depletion of high-flux, low-$\dot E$ pulsars, nor an increase of low-flux, low-$\dot E$ pulsars.
            We argue that the relatively large sample of pulsars with $\dot{E} < 10^{31} \, {\rm erg / s}$ (143 objects) and the fact that all of them have $\xi>10^{-4}$ implies that lack of detection of pulsars with low efficiency is not due to insufficient sensitivity of detectors.
            The two correlations reported above are therefore intrinsic.
            % http://127.0.0.1:8000/database/plots/radio/flux_sd (flux_sd.eps)

        \subsection{$P - \dot P$}

            There have been efforts to look for how $L$ depends on $P$ and $\dot P$ \citep[e.g.][]{1969_Ostriker,1975_Lyne,2006_Malov}.
            The results are affected by the selection effect, and have been inconclusive \cite[see review paper by][]{2013_Bagchi}.
            We will show in this section that these dependences, if any, are rather weak.

\begin{figure*}[ht]
                \begin{center}
                    \includegraphics{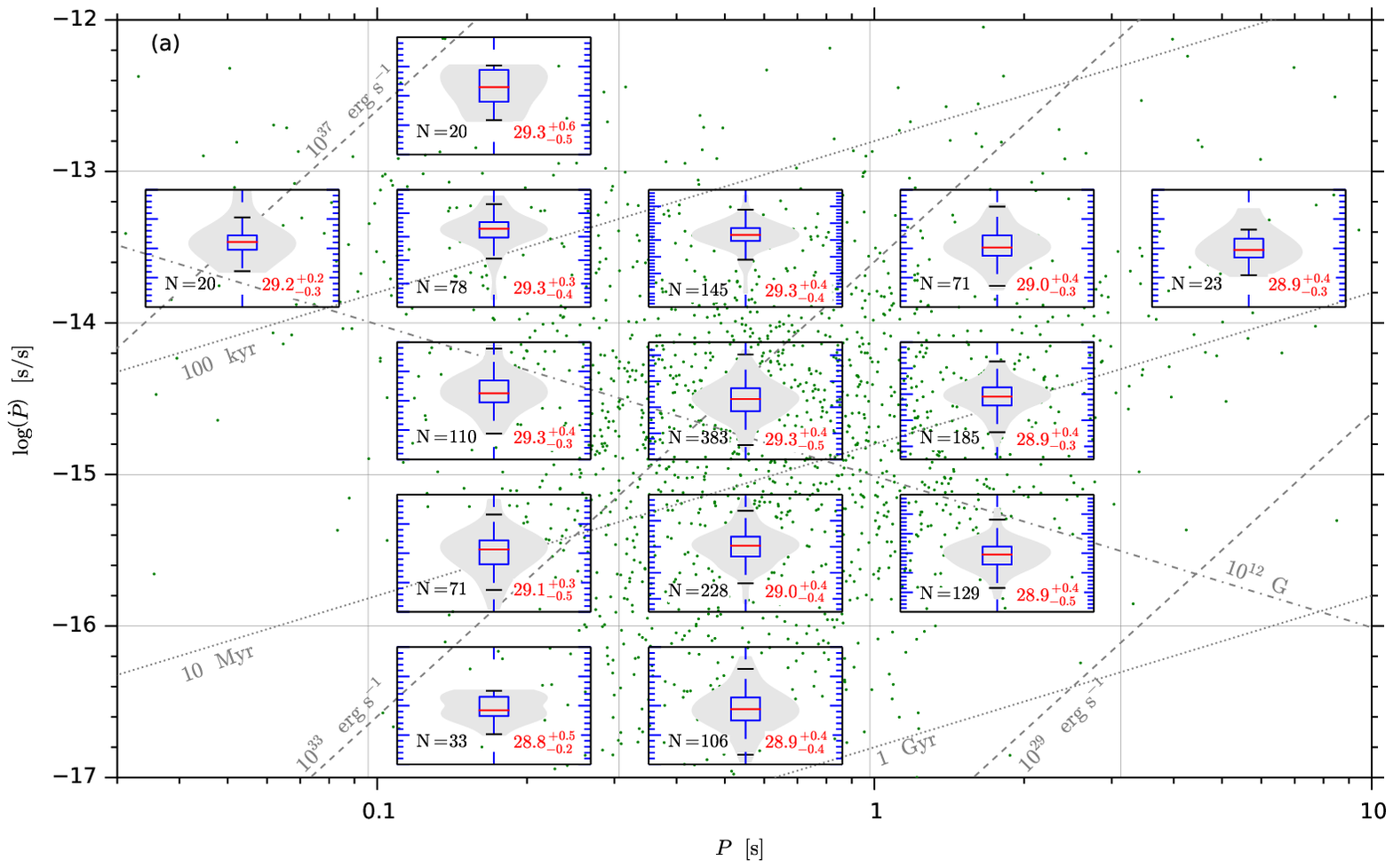}\\
                    \includegraphics{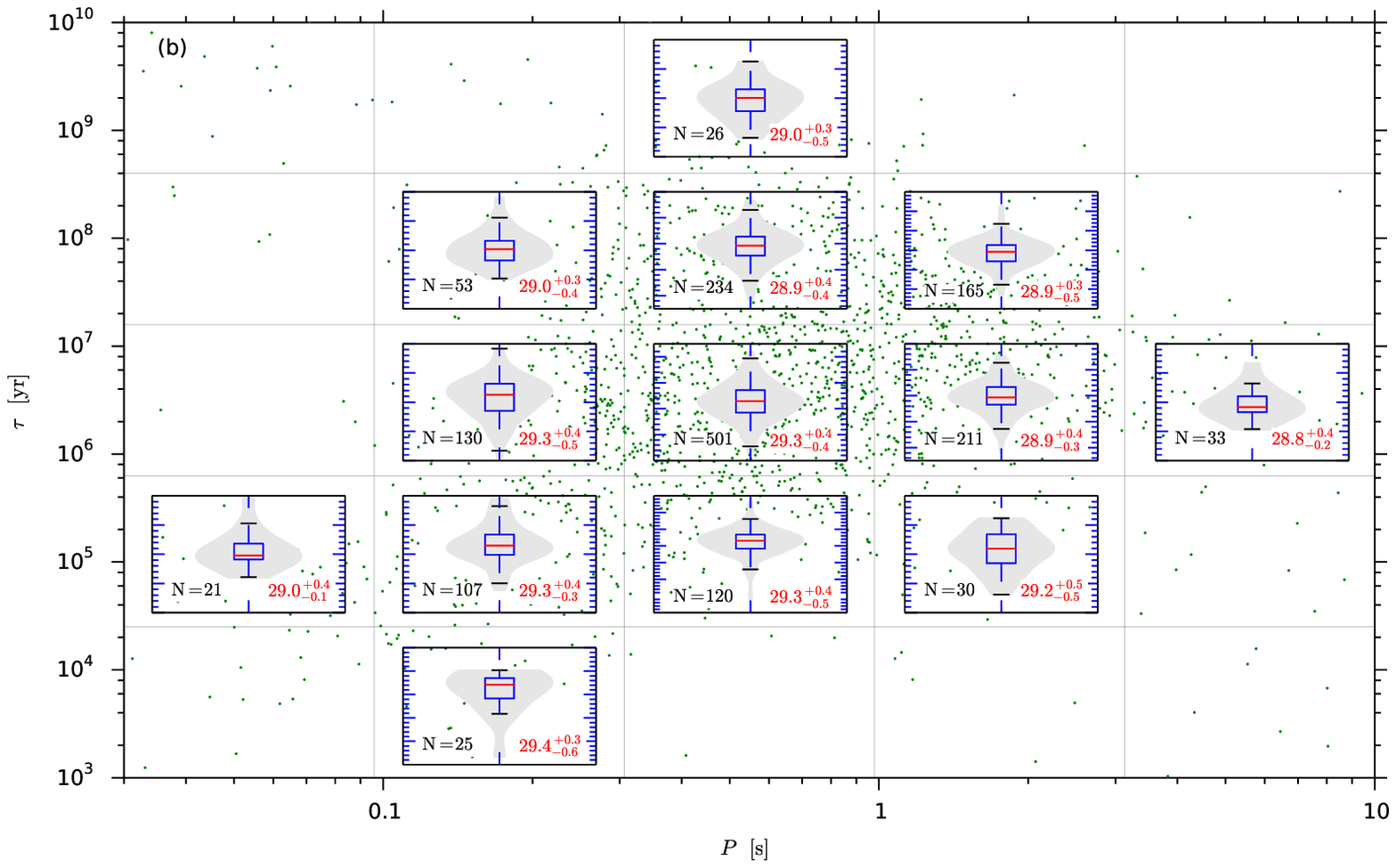}
                \end{center}
                \caption{Violin histogram plots of pulsar luminosity for the observed sample of pulsars imposed on the $P-\dot{P}$ diagram (panel a) and the $P-\tau$ diagram (panel b).
                Each histogram is calculated for a sample of pulsars restricted by a corresponding grid (gray solid lines).
                The red lines correspond to the median values of luminosity.
                The box plots show interquartile ranges, while the gray shapes represent kernel density estimations.
                In the panel (a), contours of constant spin-down luminosity, of constant characteristic age and of constant dipolar magnetic field are indicated by dashed, dotted and dot-dashed lines, respectively.
                }
                \label{fig:p_pdot_age.luminosity}
                
            \end{figure*}

            We first show the distributions of radio efficiency in different locations of the $P-\dot P$ diagram and $P-\tau$ diagram as observed (see Figure \ref{fig:p_pdot_age.luminosity}).            
            %In  Figure \ref{fig:p_pdot_age.luminosity} we show histogram plots of pulsar luminosity for the observed sample imposed on the $P-\dot{P}$ and $P-\tau$ diagrams.
            As can be clearly seen, the observed radio luminosity does not depend in any significant way on $P$ or $\dot{P}$.
            % http://127.0.0.1:8000/database/plots/radio/p_pdot/4
            % population_synthesis.py (plot_pdot 41, plot_pdot 7064)
            \begin{figure*}[ht!]
                \begin{center}
                    \includegraphics{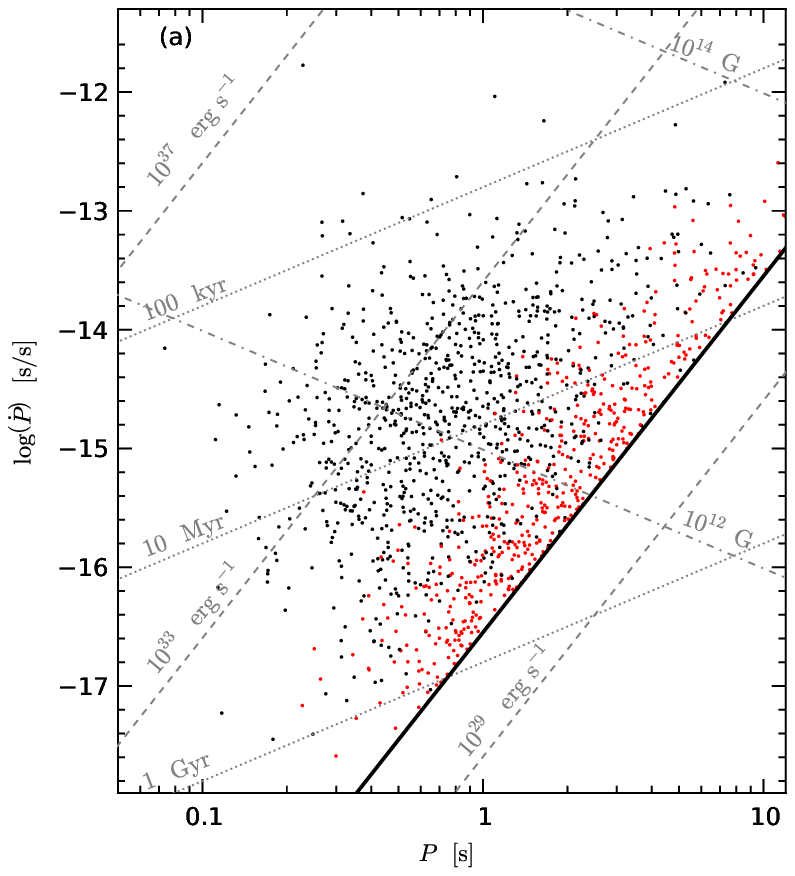}
                    \includegraphics{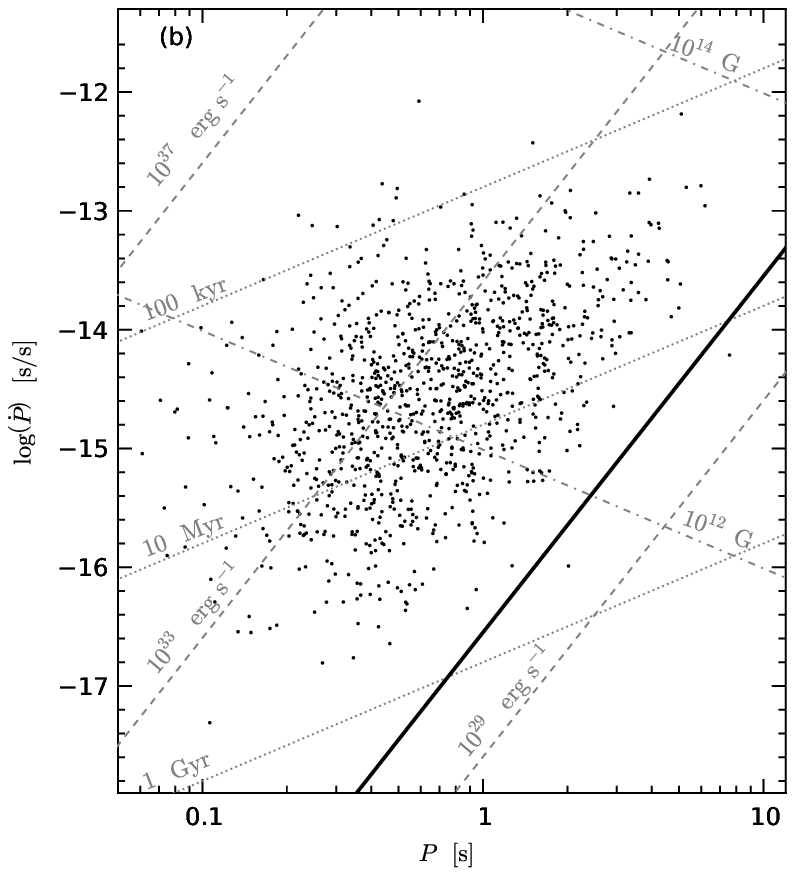}
                \end{center}
                \caption{$P-\dot{P}$ diagram for a typical MC realization calculated using parameters presented in Table \ref{tab:population_parameters} (assuming no magnetic field decay).
                Contours of constant spin-down luminosity, constant characteristic age and of constant dipolar component of magnetic field are indicated by dashed, dotted and dot-dashed lines, respectively.
                Red dots correspond to pulsars with radio efficiency greater than one percent.
                In the panel (a) the thick solid line corresponds to the modeled death line \citep{1992_Bhattacharya}, while in the MC realization presented in the panel (b) pulsars were rejected based on the radio efficiency limit $\xi < 0.01$.}
                \label{fig:faucher_random}
            \end{figure*}
            
            It would be interesting to investigate the intrinsic underlying $P$ and $\dot P$ dependences of $L$, which may be revealed through Monte Carlo simulations.
            \cite{2006_Faucher} (hereafter \citetalias{2006_Faucher}) argued that in the absence of torque decay (e.g. due to magnetic field decay) the radio luminosity of pulsars must be correlated with pulsar age, and hence with $P$ and $\dot{P}$.
            They found that the luminosity law (i.e. dependence of the radio luminosity on $P$ and $\dot{P}$) should be close to $L\propto P^{-1.5} \dot{P}^{0.5}$.
            In order to independently investigate an intrinsic luminosity distribution and eventual dependence of $L$ on $P$ and $\dot{P}$, we have performed some ``Monte Carlo'' (MC) simulations based on the open-source package \textsc{PsrPopPy} \citep{2013_Bates}.
            This package includes two methods to obtain a synthetic sample of pulsars: the snapshot and evolutionary methods.
            The snapshot method consists of the following steps: generating pulsar periods, modeling of pulse widths, generating radio luminosities, distributing the pulsars in the Galaxy, modeling electron density, and generating spectral indices.
            The evolutionary method consists of the steps from the snapshot method but is also extended by additional ones: generating pulsar period derivatives, generating magnetic fields, generating rotational alignment and modeling its time evolution, modeling pulsar spindown, and finally, evolving pulsars through the Galactic potential  \citep[see][for more details]{2013_Bates}.
            Therefore, we used the evolutionary part of this code to reproduce the results of \citetalias{2006_Faucher}.
            After the calculation of a synthetic population, each pulsar is run through parameters of the {\it Parkes Multibeam Pulsar Survey} \citep{2001_Manchester} (PMPS) in order to constrain the population based upon known detections.
            PMPS is the most successful survey to date with 1062 normal pulsars detected, which allows one to test a synthetic population using more than a half of the total observed population of normal pulsars.
            
                \begin{table}[h!]
                    \caption{Population synthesis parameters used in MC simulations based on the evolutionary method.}
                    \begin{center}
                        {\tiny
                        \begin{tabular}{p{4.2cm}r}
                            \hline
                            \hline
                            & \\
                            Radial distribution model & \cite{2004_Yusifov} \\
                            $R_1$ \dotfill & $55 \,{\rm pc}$ \\
                            $a$ \dotfill & $1.64$ \\
                            $b$ \dotfill & $4.01$ \\                           
                            %Radial distribution model & \cite{2006_Lorimer} \\
                            %$a$ \dotfill & $3.51$ \\
                            %$b$ \dotfill & $7.89$ \\                           
                            Birth height distribution & Exponential \\
                            Initial Galactic z-scale height \dotfill & $50 \,{\rm pc}$ \\
                            %Birth velocity distribution & Gaussian \\
                            %$\left < v_{x,y,z}\right >$ & $180 \, {\rm km\,s^{-1}}$ \\
                            Birth velocity distribution & Exponential \\
                            $\left < v_{3\rm D}\right >$ & $380 \, {\rm km\,s^{-1}}$ \\
                            & \\
                            Pulsar spin-down model & \cite{2006_Faucher} \\
                            Beam alignment model & Orthogonal \\
                            Breaking index & $3.0$ \\
                            %\hline
                            Birth spin period distribution & Normal \\
                            $\left < P_{0} \right >$ \dotfill & $300 \,{\rm ms}$ $^{a}$, $200 \,{\rm ms}$ $^{b}$\\
                            $\sigma_{P_0}$ \dotfill & $150 \,{\rm ms}$ $^{a, b}$\\
                            %\hline
                            %Birth spin period distribution (optimal) & Normal \\
                            %$\left < P_{0} \right >$ \dotfill & $100 \,{\rm ms}$ \\
                            %$\sigma_{P_0}$ \dotfill & $200 \,{\rm ms}$ \\                           
                            %\hline
                            & \\
                            Scattering model & \cite{2004_Bhat}\\
                            Spectral index of the scattering model & $-3.86$\\
                            Maximum age of pulsars \dotfill & $1 \, {\rm Gyr}$ \\
                            Spectral index distribution model & Normal\\
                            $\left <\alpha \right >$ \dotfill & $-1.4$ \\
                            $\sigma_{\alpha}$ \dotfill & $0.96$ \\
                            & \\
                            Magnetic field distribution model & Lognormal \\
                            $\left <\log_{10} \left ( B_{\rm d} [{\rm G}] \right )  \right >$ \dotfill & $12.65$ \\
                            $\sigma_{\log_{10} B_{\rm d} }$ \dotfill & $0.55$ \\
                            %& \\
                            %\hline
                            & \\
                            Random luminosity distribution model$^a$ & Lognormal \\
                            $\left <\log_{10} \left ( L_{1400} [{\rm mJy\,kpc^2}] \right )  \right >$ \dotfill & $-1.1$ $^{a}$, $0.5$ $^{b}$\\
                            $\sigma_{\log_{10} L_{1400}}$ \dotfill & $0.9$ $^{a}$, $1.0$ $^{b}$ \\
                            & \\
                            \hline
                            %& \\
                            %Random luminosity distribution model (optimal)$^b$ & Lognormal \\
                            %$\left <\log_{10} \left ( L_{1400} [{\rm mJy\,kpc^2}] \right )  \right >$ \dotfill & $0.63$ \\
                            %$\sigma_{\log_{10} L_{1400}}$ \dotfill & $0.74$ \\                           
                            %Power-law luminosity model$^b$ &  $\displaystyle L_{1400}=L_0 P^p \dot{P}^q + 10^{L_{\rm c}}$\\
                            %$L_0$ \dotfill & $0.18\,{\rm mJy\,kpc^2}$ \\
                            %$p$ \dotfill & $-1.5$ \\
                            %$q$ \dotfill & $0.5$ \\
                            %$\sigma L_{\rm c}$ \dotfill & $0.8$ \\
                            %& \\
                            \hline
                            %& \\
                            Number of pulsars detected in PMPS & 1100 \\
                            %& \\
                            \hline
                            \hline
                        \end{tabular}
                        }
                    \end{center}
                    \label{tab:population_parameters}
                    {\tiny
                    $^{a}$ - population model parameters used to reproduce results of FK06\\
                    $^{b}$ - optimal population model parameters
                    }
                \end{table}
                
               \newpage 
               
            Table \ref{tab:population_parameters} summarizes the parameters used by \citetalias{2006_Faucher} to produce their Figures 7 and 14 calculated, as well as the best fit parameters we have obtained from our simulations (see more details below).

            Figure \ref{fig:faucher_random}a shows the reproduced $P-\dot{P}$ diagram for the random luminosity model calculated using the theoretical death line approximated by the equation \mbox{$B_{\rm d} / P^2 =0.17\times 10^{12} \,{\rm G\,s^{-2}}$} \citep{1992_Bhattacharya} (Figure 14 in \citetalias{2006_Faucher}).
            The result is similar to \citetalias{2006_Faucher}, who noted that, in the absence of magnetic field decay and using the random luminosity model, the results of Monte Carlo-based population synthesis showed a clear pileup of observed objects near the death line in the $P-\dot{P}$ diagram.
            However, we note that the used approach did not take into account that a radio emission process (whatever it may be) should have an upper limit for its efficiency.
            In Figure \ref{fig:faucher_random}a pulsars with derived radio efficiency greater than one percent are marked by reds dots.
            The efficiency of some of these pulsars approaches, and even exceeds 100\%.
            This is physically unreasonable.
            More likely, the condition for pulsar radio emission may be such that a certain maximum efficiency is imposed. 
%            We believe that the pileup of pulsars near the death line reported by \citetalias{2006_Faucher} was a direct consequence of using incomplete (inadequate) condition for pulsar death.
            One should check whether the radio efficiency (see Equations \ref{eq:eff} and \ref{eq:lum}) does not exceed 100\% ($\xi<1$) for each independent draw of $L_{1400}$, $P$ and $B_{\rm d}$.
            
            In Figure \ref{fig:faucher_random}b we show the $P-\dot{P}$ diagram obtained by MC simulation using our optimal random luminosity model (see Table \ref{tab:population_parameters}) with the condition for pulsar death based only on the upper limit for radio efficiency \mbox{$\xi_{\rm max} =0.01$}.
            Using the efficiency limit allows not only to avoid a pileup of pulsars near the theoretical death line, but also explains the existence of a few pulsars observed in the so-called graveyard region.

            % population_synthesis / plot_sets  3500-3549, 3700-3749, 1100-1149
            \begin{figure*}[ht!]
                \begin{center}
                    \includegraphics{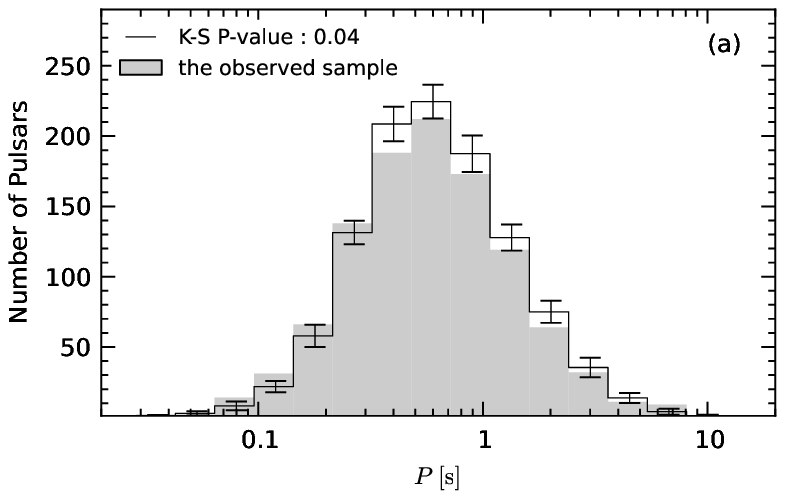}
                    \includegraphics{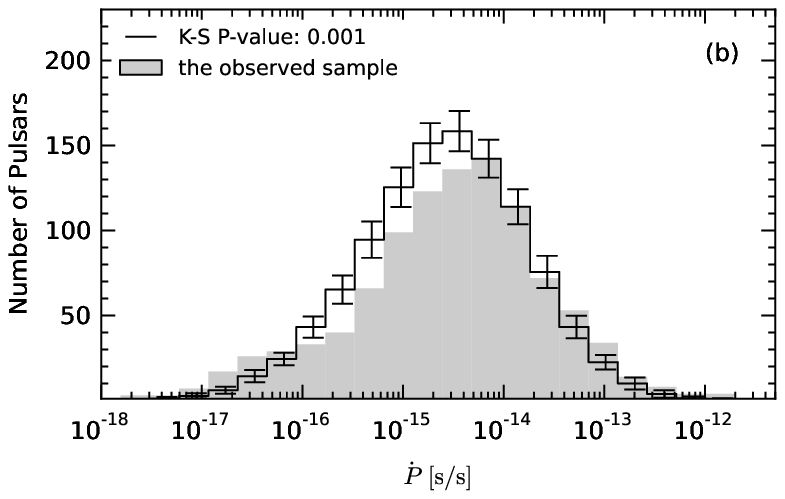}\\
                    \includegraphics{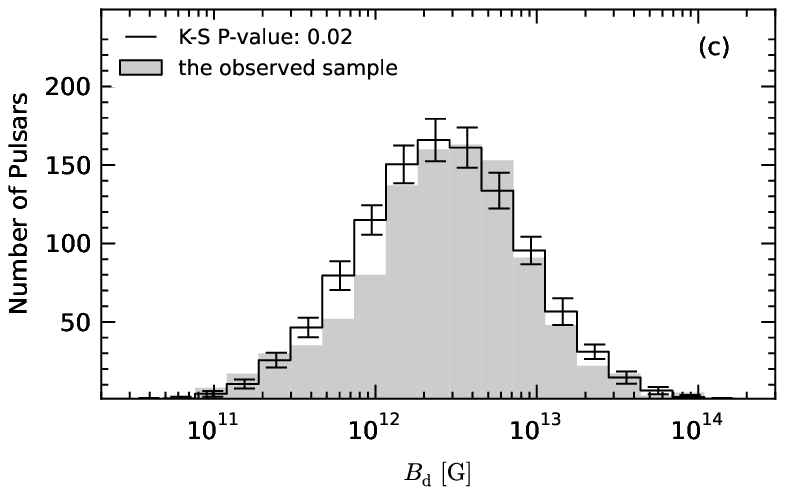}
                    \includegraphics{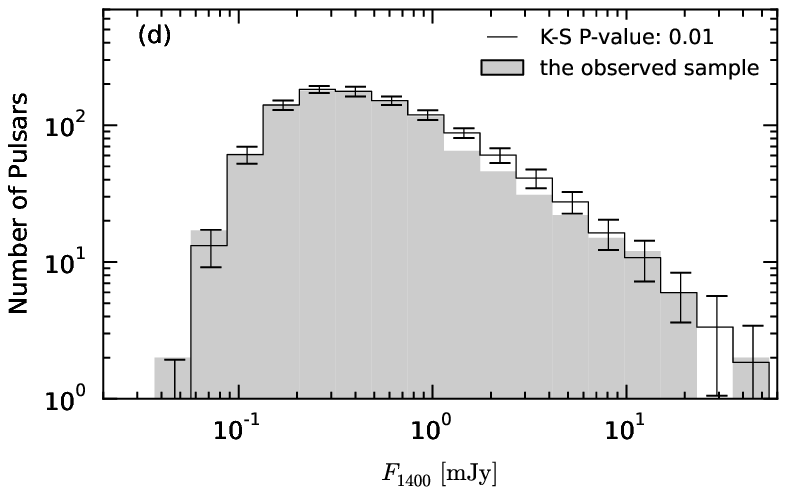}
                \end{center}
                \caption{
                Distributions of pulse periods (panel a), period derivatives (panel b), dipolar components of magnetic field at the polar cap (panel c) and radio flux (panel d) compared to the real distributions (filled gray histograms). 
                Distributions correspond to MC simulations using the optimal random luminosity model (see Table \ref{tab:population_parameters}).
                For histograms of synthetic data, the number of pulsars in each bin is the average over $50$ MC realizations and the error bars indicate the corresponding standard deviations.
                On each histogram, the corresponding probability from the K-S test is displayed in the legend. All MC pulsars ($50\times 1100$) were used to perform the K-S tests.
                %We present both, the result from the K-S test for all pulsars from 50 MC realizations and the maximum value of the probability found for a single MC realization.
                }
                \label{fig:histograms}
            \end{figure*}

            In Figure \ref{fig:histograms} we present comparison of the observed distribution with synthetic distributions for our optimal model (presented in Figure \ref{fig:faucher_random}b).
            We have found that the random luminosity model with proper input parameters can reproduce the observed sample much better than the models proposed by FK06.
            The Kolmogorov-Smirnov (K-S) test on the flux distribution for the power-law model gives the probability (the K-S P-value) $10^{-8}$ (see Figure 6 in \citetalias{2006_Faucher}), while the K-S test on the flux distribution for our optimal random luminosity model result in probability of $1\%$.
            Note that in our optimal model the distribution of the spin period at pulsar birth is centered at $200\,{\rm ms}$.
            This allowed to obtain the higher probability of reproducing the period distribution than the one obtained by FK06 (compare $4\%$ with $0.7\%$), but negatively affected the probability of reproducing the period derivative distribution ($0.1\%$), and hence, the $B_{\rm d}$ distribution (compare $2\%$ with $15\%$).
            It is worth noting that one of the parameters which strongly affects both period and period derivative distributions is the maximum efficiency limit.
            Taking into account a relatively large number of parameters required by the evolutionary method and the fact that we arbitrary choose \mbox{$\xi_{\rm max} = 0.01$}, the parameters of the model can be further optimized in order to better reproduce the observed sample.
            Calculations show that using the condition for pulsar death based on the radio efficiency limit allows to avoid pile up of pulsars near the death line.
            As a result we have found that neither the magnetic field decay nor the dependence of the luminosity on $P$ and $\dot{P}$ are required to reproduce the observed sample\footnote
{            This does not mean that the magnetic field decay is not significant over lifetime of pulsars as radio-loud sources.
            We note that even if there is a significant decay of the magnetic filed it should be related to a change of the global (dipolar) magnetic field.
            On the other hand, the existence of non-dipolar configuration of the surface magnetic field was postulated since the very beginning of radio astronomy \citep{1975_Ruderman}.
            We believe that parameters of the plasma (which is responsible for radio emission) strongly depend on the curvature and strength of surface magnetic field.
            The fact that radio luminosity does not depend in any significant way on the dipolar component of magnetic field ($B_{\rm d}$) is a strong proof that the magnetic field at stellar surface is dominated by crust-anchored local magnetic anomalies (see Section \ref{sec:sum} for more details).}.
            
\begin{figure}[ht!]
                \begin{center}
                    \includegraphics[width=8.1cm]{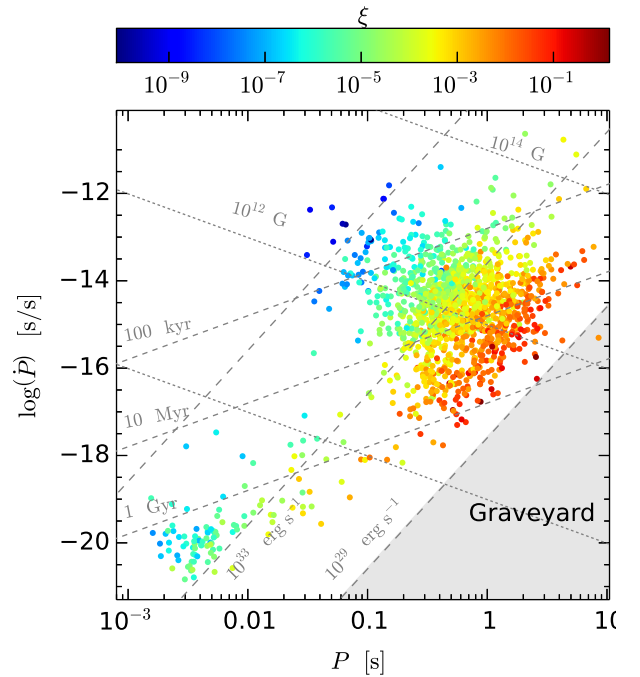}
                \end{center}
                \caption{
                The $P-\dot{P}$ diagram shown for a sample 1436 pulsars used in the analysis of radio efficiency.
                Colors correspond to different values of radio efficiency.
                }
                \label{fig:p_pdot_efficiency}
            \end{figure}

            Figure \ref{fig:p_pdot_efficiency} presents the $P-\dot{P}$ diagram for the whole sample of observed pulsars studied in this paper, with the color scheme denoting how radio efficiency $\xi$ is distributed.
            The figure clearly shows that the radio emission graveyard in the $P -\dot P$ diagram corresponds to a region that $\xi$ may exceed its upper limit (a few percent).
            
            \hspace{0.5cm} % ugly hack here...
            %As can be clearly seen, the radio efficiency (and thereby also luminosity) does not depend in any significant way on $P$ or $\dot{P}$.
           %On the contrary, the most noticeable dependence is the anticorrelation between $\xi$ and $\dot{E}$ and a positive dependence between $\xi$ and $\tau$.

            %http://127.0.0.1:8000/database/plots/radio/p_pdot/3

    \section{Summary and discussion} \label{sec:sum}
        Using a large sample of pulsars from the ATNF catalog, we draw the following conclusions in this paper: 1. Radio pulsar luminosity has a very weak, if any, dependence on pulsar spin-down luminosity $\dot E$ over near 8 orders of magnitude in $\dot E$.
        There might be a slight variation of the exponent of the $L-\dot E$ correlation with frequency.
        2. There is a near linear inverse correlation between radio efficiency $\xi$ and $\dot E$ for both normal and binary/millisecond pulsars.
        3. The radio efficiency $\xi$ is roughly linearly correlated with pulsar age $\tau$ for normal pulsars.
        4. The two reported correlations are not due to an observational selection effect.
        5. Since radio luminosity does not depend on $P$ and $\dot{P}$ (or the dependence is very weak) the proposed mechanism of radio emission must explain the significant increase of radio efficiency for low-$\dot{E}$ pulsars.

        A weak $L-\dot E$ correlation was hinted by previous studies \citep[e.g.][]{1993_Lorimer,2010_Ridley}.
        \cite{2006_Malov} found a weak positive dependence $L \propto \dot{E}^{0.29}$ with a smaller sample (338 objects), which was selected by requiring the pulsars to have well-determined spectra, and/or well-determined distance.
        We adopt a much larger sample, which has the following advantages. First, a larger sample size can insure more reliable correlations. Second, many uncertainties, e.g. unknown moment of inertia of neutron stars, the influence of scintillation on pulsar flux density, or how well the one-dimensional line-of-sight cut through an emission beam could represent the entire beam, can be effectively canceled out.
        With our enlarged sample, the index of the correlation, if any, is much flatter than the ones found in the previous studies \cite[see e.g.][]{2002_Arzoumanian, 2006_Malov}.

        As discussed in \S2.1, by using the simplified formula (\ref{eq:lum}), we have neglected some complicated pulsar geometry factors, such as $\alpha$, $\beta$ (which defines the duty cycle $\delta$ along with $P$, and $\rho$).
        For individual pulsars, the derived $L$ should have a large error.
        Including the entire sample would cancel out most of these geometric factors, so that our discovered $\xi-\dot E$ and $\xi-\tau$ correlations would be intrinsic.
        One possible factor that is not fully canceled is the $P$-dependence.
        This is because both $\rho$ and $\delta$ depend on $P$.
        However, the explicit exponent of $P$-dependence is unknown.
        If we assume that the mechanism of radio emission is the same for both normal and millisecond pulsars, we find that by adding a dependence of $\sim P^{-0.5}$ to the integrated luminosity, the $\xi-\dot E$ correlations for both normal and binary/millisecond pulsars become consistent with each other.
        Adding this period dependence results in the following luminosity relationships: $L_{400}\propto \dot{E}^{0.27}$, $L_{1400}\propto \dot{E}^{0.19}$, $L_{2000}\propto \dot{E}^0$.
        The found correlations are closer but still shallower than the one found by \cite{2006_Malov}.
        With such a correction, the $\xi-\dot E$ correlation is slightly shallower, i.e. $\xi \propto \dot E^{-0.81}$, but is still very significant.
        It has an even lesser impact on the $\xi-\tau$ correlation, i.e. $\xi \propto \tau^{1.01}$ after correction.
        In a more general form, we can write that  $\delta^{-1} \sin^2{(\rho/2)} \propto P^p \dot{P}^q $.
        Since in our analysis we are not using geometry information we cannot unambiguously define both $p$ and $q$.
        However, assuming similar radio luminosity of  normal and millisecond pulsars we can write that $ \delta^{-1} \sin^2{(\rho/2)} \propto P^{p} \dot{P}^{-0.16(2p+1)}$.
        Note that even introducing some model-dependent value of $p$ does not change the general picture presented in this paper.

        It would seem natural to assume that the spin-down parameters affect the radio luminosity (and hence the radio efficiency) of pulsars.
        Therefore, many authors \citep[e.g.][]{1969_Ostriker,1975_Lyne, 1986_Stollman, 2006_Malov, 2006_Faucher} tried to define the luminosity law, however, the results vary greatly (depending on pulsar samples and used methods) and were inconclusive \cite[see e.g.][]{2013_Bagchi}.
        The spin-down parameters determine the magnetic field strength at the light cylinder \mbox{$B_{\rm LC} \propto P^{-2.5} \dot{P}^{0.5}$}.
        Then, assuming dipolar configuration of magnetic field we can estimate the field strength at the stellar surface as \mbox{$B_{\rm d} \propto (P \dot{P})^{0.5}$} and the vacuum potential drop as \mbox{$V\propto B_{\rm d} P^{-1} \propto P^{-0.5} \dot{P}^{0.5}$} \citep{1975_Ruderman}.
        The main parameters that affects properties of pulsar radio emission are density and distribution of the electron-positron pair plasma.
        To estimate the plasma properties we need to use some specific model of acceleration (e.g. the vacuum gap, the space-charged limited flow, the partially screened gap models) and also to take into account various emission processes (e.g. curvature radiation, inverse Compton scattering).
        It was shown \citep[see e.g.][]{2000_Zhang_a,2001_Hibschman, 2001_Hibschman_b} that the plasma density and distribution highly depend on some factors which cannot be estimated by the spin-down parameters.
        %Estimation of these parameters is not possible taking into account only radio observations of pulsars.
        Thus, we cannot specify the properties of plasma responsible for radio emission using only $P$ and $\dot{P}$.
        However, assuming dipolar configuration of pulsar magnetic field, some theoretical predictions suggest that the older the pulsar, the smaller the final pair multiplicity \citep{2001_Hibschman, 2001_Hibschman_b}.
        It would suggest that some kind of dependence of the radio luminosity on the pulsar age could also be found.
        As we have shown it is not the case for the observed sample of radio pulsars (see Figure \ref{fig:p_pdot_age.luminosity}).
        
        In order to reproduce the observed sample of pulsars FK06 argued that in the absence of torque decay pulsar radio luminosity has to depend on $P$ and $\dot{P}$ as \mbox{$L\propto P^{-1.5}\dot{P}^{0.5}$} (or \mbox{$L\propto P^{-1.39}\dot{P}^{0.48}$} with exponents optimized by \citet{2013_Bates}) . 
        However, the power-law luminosity model results in relatively low P-value of the K-S test for the flux distribution.
        We have found that replacing the modeled death line \citep{1992_Bhattacharya} by the radio efficiency limit allows to reproduce the observed sample even with the luminosity model previously excluded by FK06, namely the random luminosity model.
        Introducing the radio efficiency limit into MC simulations results in much higher probabilities from the K-S test, allows to avoid a pileup of pulsars near the theoretical death line and, furthermore, explains the existence of a few pulsars observed in the so-called graveyard region.
        We argue that the facts that the observed luminosity does not depend on both $P$ and $\dot{P}$, and that the random luminosity model of the intrinsic luminosity allows to reproduce the observed sample is a strong indication in favor of the random model.
        
        It is very difficult to explain the spread of up to four orders of magnitude for both $L$ and $\xi$ for a given spin-down luminosity only by statistical uncertainties.
        As we have mentioned above there must be some parameter/parameters other than $P$, $\dot P$ that influence the radio efficiency of pulsars.
        The clear $\xi-\dot E$ linear inverse correlation and $\xi - \tau$ linear correlation for radio emission call for the following physical picture: The plasma responsible for generating pulsar radio emission must be produced under similar conditions regardless of pulsar age, dipolar magnetic field strength, and spin down rate.
        One possibility is that the pulsar polar cap region may have dominant crust-anchored magnetic anomalies, so that the near surface magnetic field configuration significantly deviates from the dipolar geometry \citep{2002_Gil}.
        X-ray observations of old pulsars indeed revealed hot spots that are significantly smaller than the conventional polar cap \citep{2005_Zhang,2008_Gil,2013_Szary}, which is consistent with having strong non-dipolar magnetic field at the surface due to crust-anchored local anomalies.% \citep{2003_Geppert}.

\begin{figure}[h!]
                \begin{center}
                    \includegraphics{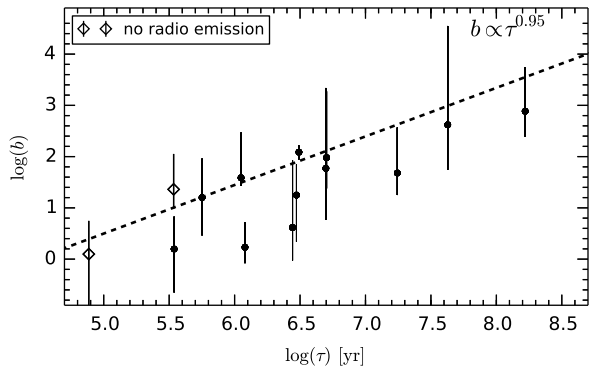}
                \end{center}
                \caption{Dependence of the $b$ parameter on pulsar age $\tau$ (see text for more details).
                Dashed line correspond to the linear fit for all pulsars with detected radio emission.
                Note that for \mbox{PSR J0108-1431} we used the single blackbody fit performed by \cite{2009_Pavlov}.}
                \label{fig:bparameter}
            \end{figure}

        The crust-anchored anomalies can be characterized by a parameter $b=B_{\rm s}/B_{\rm d}=A_{\rm dp}/A_{\rm bb}$, which describes the ratio of the actual value of the magnetic field at the surface $B_{\rm s}$ to the dipolar component of the magnetic field $B_{\rm d}$ \cite[see e.g.][]{2002_Gil_b}.
        Here \mbox{$A_{\rm dp}\approx 6.2 \times 10^4 P^{-1} \, {\rm m^2}$} is the conventional polar cap area (i.e. calculated assuming pure dipolar configuration of the magnetic field), and $A_{\rm bb}$ is the actual hot spot area (i.e. derived from X-ray observations).
        The sample of pulsars with an X-ray hot spot for which $b$ can be estimated is small. Nevertheless, with this small sample \citep{2013_Szary}, we can clearly see a correlation between pulsar age and $b$ (see Figure \ref{fig:bparameter}).
        
             % b_age.eps
        
        The found correlation indicates that when a pulsar becomes older, its surface magnetic field becomes more dominated by the crust-anchored magnetic anomalies generated e.g. by the Hall drift \cite[see][]{2013_Geppert, 2013_Vigano}.
        Such a crust-anchored anomaly does not depend on $\dot E$ and age, which allows pulsars to produce enough electron-positron pairs at an old age to power radio emission.

        The fact that pulsars near the graveyard tend to have a very high $\xi$ also has a profound implications in understanding radio pulsar death.
        Traditionally, pulsar death line \citep{1975_Ruderman,1993_Chen} was defined by the condition of production of electron-positron pairs, which depends on many uncertainties, including the near-surface $\gamma$-ray emission mechanism \citep{2000_Zhang} and the magnetic field configurations \citep{2001_Gil}.
        The results presented in this paper offer a simpler interpretation to radio pulsar death.
        It is possible that the required physical condition to power radio emission is similar among pulsars in all ages, which requires a certain minimum spin-down power.
        When pulsars spin down slightly below this threshold, the radio emission mechanism simply cannot operate.
        This explains why high $\xi$ pulsars are located near the death line.
        A similar conclusion was drawn by \cite{2002_Arzoumanian} based on modelling ``of the birth properties and rotational, kinematic, and luminosity evolution of a Monte Carlo population of neutron stars'', even though the luminosity and spin-down rate scaling presented in their work, $L \propto \dot{E}^{0.5}$, does not reflect the actual $L-\dot{E}$ relationship (see Figure \ref{fig:one}a).

    \acknowledgments
        This work is partially supported by a visitor program of Kavli Institute for Astronomy and Astrophysics, Peking University, China, and National Science Centre Poland under grants 2011/03/N/ST9/00669 and DEC-2012/05/B/ST9/03924.
        B.Z. acknowledges NASA NNX10AD48G for support, and R.X.X. acknowledges support by the National Science Foundation of China under Grant No. 11225314.
        We thank the anonymous referee for constructive comments.
        %\clearpage


\begin{thebibliography}{}
        \bibitem[Arons(1996)]{1996_Arons} Arons, J.\ 1996, \aaps, 120, 49

        \bibitem[Arzoumanian et al.(2002)]{2002_Arzoumanian} Arzoumanian, Z., Chernoff, D.~F., \& Cordes, J.~M.\ 2002, \apj, 568, 289

        \bibitem[Bagchi(2013)]{2013_Bagchi} Bagchi, M.\ 2013, arXiv:1306.2152

        \bibitem[Bates et al.(2013)]{2013_Bates} Bates, S., Lorimer, D., Rane, A., \& Swiggum, J.\ 2013, arXiv:1311.3427

        \bibitem[Becker(2009)]{2009_Becker} Becker, W.\ 2009, Astrophysics and Space Science Library, 357, 91

        \bibitem[Bhat et al.(2004)]{2004_Bhat} Bhat, N.~D.~R., Cordes, J.~M., Camilo, F., Nice, D.~J., \& Lorimer, D.~R.\ 2004, \apj, 605, 759

       \bibitem[Bhattacharya et al.(1992)]{1992_Bhattacharya} Bhattacharya, D., Wijers, R.~A.~M.~J., Hartman, J.~W., \& Verbunt, F.\ 1992, \aap, 254, 198

        \bibitem[Chen \& Ruderman(1993)]{1993_Chen} Chen, K., \& Ruderman, M.\ 1993, \apj, 402, 264

        \bibitem[Contopoulos \& Spitkovsky(2006)]{2006_Contopoulos} Contopoulos, I., \& Spitkovsky, A.\ 2006, \apj, 643, 1139

        \bibitem[Faucher-Gigu{\`e}re \& Kaspi(2006)]{2006_Faucher} Faucher-Gigu{\`e}re, C.-A., \& Kaspi, V.~M.\ 2006, \apj, 643, 332, \citetalias{2006_Faucher}

        %\bibitem[Geppert et al.(2012)]{2012_Geppert} Geppert, U., Gil, J., Melikidze, G., Pons, J., \& Vigan{\`o}, D.\ 2012, Astronomical Society of the Pacific Conference Series, 466, 187

        \bibitem[Geppert et al.(2013)]{2013_Geppert} Geppert, U., Gil, J., \& Melikidze, G.\ 2013, \mnras, 435, 3262

        %\bibitem[Geppert et al.(2003)]{2003_Geppert} Geppert, U., Rheinhardt, M., \& Gil, J.\ 2003, \aap, 412, L33

        \bibitem[Gil \& Mitra(2001)]{2001_Gil} Gil, J., \& Mitra, D.\ 2001, \apj, 550, 383

        \bibitem[Gil et al.(2002a)]{2002_Gil} Gil, J.~A., Melikidze, G.~I., \& Mitra, D.\ 2002a, \aap, 388, 235

        \bibitem[Gil et al.(2002b)]{2002_Gil_b} Gil, J.~A., Melikidze, G.~I., \& Mitra, D.\ 2002b, \aap, 388, 246

        \bibitem[Gil et al.(2008)]{2008_Gil} Gil, J.~A. et al. 2008, \apj, 686, 497

        \bibitem[Gunn \& Ostriker(1970)]{1970_Gunn} Gunn, J.~E., \& Ostriker, J.~P.\ 1970, \apj, 160, 979


        \bibitem[Hibschman \& Arons(2001a)]{2001_Hibschman} Hibschman, J.~A., \& Arons, J.\ 2001a, \apj, 560, 871

        \bibitem[Hibschman \& Arons(2001b)]{2001_Hibschman_b} Hibschman, J.~A., \& Arons, J.\ 2001b, \apj, 554, 624
        
        \bibitem[Lorimer et al.(1993)]{1993_Lorimer} Lorimer, D.~R., Bailes, M., Dewey, R.~J., \& Harrison, P.~A.\ 1993, \mnras, 263, 403

        \bibitem[Lorimer et al.(2004)]{2004_Lorimer} Lorimer, D.~R., Kramer, M., Ellis, R., et al.\ 2004, Handbook of pulsar astronomy, by D.R.~Lorimer and M.~Kramer.~Cambridge observing handbooks for research astronomers, Vol.~4.~Cambridge, UK: Cambridge University Press, 2004,

        \bibitem[Lorimer et al.(2006)]{2006_Lorimer} Lorimer, D.~R., Faulkner, A.~J., Lyne, A.~G., et al.\ 2006, \mnras, 372, 777

        %\bibitem[Lorimer \& Kramer(2012)]{2012_Lorimer} Lorimer, D.~R., \& Kramer, M.\ 2012, Handbook of Pulsar Astronomy, by D.~R.~Lorimer , M.~Kramer, Cambridge, UK: Cambridge University Press, 2012,

        \bibitem[Lyne et al.(1975)]{1975_Lyne} Lyne, A.~G., Ritchings, R.~T., \& Smith, F.~G.\ 1975, \mnras, 171, 579

        \bibitem[Malov \& Malov(2006)]{2006_Malov} Malov, I.~F., \& Malov, O.~I.\ 2006, Astronomy Reports, 50, 483

        \bibitem[Malov \& Malov(2007)]{2007_Malov} Malov, I.~F., \& Malov, O.~I.\ 2007, VizieR Online Data Catalog, 908, 30542

        \bibitem[Manchester et al.(2005)]{2005_Manchester} Manchester, R.~N., Hobbs, G.~B., Teoh, A., \& Hobbs, M.\ 2005, \aj, 129, 1993

        \bibitem[Manchester et al.(2001)]{2001_Manchester} Manchester, R.~N., Lyne, A.~G., Camilo, F., et al.\ 2001, \mnras, 328, 17

        \bibitem[Mastrano et al.(2013)]{2013_Mastrano} Mastrano, A., Lasky, P.~D., \& Melatos, A.\ 2013, arXiv:1306.4503

%        \bibitem[McLaughlin et al.(2003)]{2003_McLaughlin} McLaughlin, M.~A., Stairs, I.~H., Kaspi, V.~M., et al.\ 2003, \apjl, 591, L135

        \bibitem[Melrose(2006)]{2006_Melrose} Melrose, D.~B.\ 2006, Chinese Journal of Astronomy and Astrophysics Supplement, 6, 020000

        \bibitem[Ostriker \& Gunn(1969)]{1969_Ostriker} Ostriker, J.~P., \& Gunn, J.~E.\ 1969, \nat, 223, 813

        \bibitem[Pavlov et al.(2009)]{2009_Pavlov} Pavlov, G.~G., Kargaltsev, O., Wong, J.~A., \& Garmire, G.~P.\ 2009, \apj, 691, 458

        \bibitem[Ridley \& Lorimer(2010)]{2010_Ridley} Ridley, J.~P., \& Lorimer, D.~R.\ 2010, \mnras, 404, 1081

        \bibitem[Ruderman \& Sutherland(1975)]{1975_Ruderman} Ruderman, M. A., \& Sutherland, P. G. 1975, \apj, 196, 51

        \bibitem[Stollman(1986)]{1986_Stollman} Stollman, G.~M.\ 1986, \aap, 170, 48


        \bibitem[Szary(2013)]{2013_Szary} Szary, A.\ 2013, arXiv:1304.4203, PhD Thesis

        \bibitem[Taylor \& Manchester(1977)]{1977_Taylor} Taylor, J.~H., \& Manchester, R.~N.\ 1977, \apj, 215, 885

        \bibitem[The Fermi-LAT collaboration(2013)]{2013_Fermi} The Fermi-LAT collaboration 2013, arXiv:1305.4385
        
        \bibitem[Vigan{\`o} et al.(2013)]{2013_Vigano} Vigan{\`o}, D., Rea, N., Pons, J.~A., et al.\ 2013, \mnras, 434, 123 
        
%       \bibitem[Young et al.(1999)]{1999_Young} Young, M.~D., Manchester, R.~N., \& Johnston, S.\ 1999, \nat, 400, 848
        \bibitem[Yusifov \& Kucuk (2004)]{2004_Yusifov} Yusifov, I., Kucuk, I.\ 2004, \aap, 422, 545
        
        \bibitem[Zhang \& Harding(2000)]{2000_Zhang_a} Zhang, B., \& Harding, A.~K.\ 2000, \apj, 532, 1150

        \bibitem[Zhang et al.(2000)]{2000_Zhang} Zhang, B., Harding, A. K., \& Muslimov, A. G. 2000, \apj, 531, L135


        \bibitem[Zhang et al.(2005)]{2005_Zhang} Zhang, B., Sanwal, D., \& Pavlov, G. G. 2005, \apj, 624, L109
    \end{thebibliography}
\end{document}